\documentclass[11pt]{article}
\usepackage{moriond,epsfig}
\usepackage{amssymb}
\usepackage{dcolumn}
\usepackage{graphicx}
\usepackage{amsmath}

\def\be{\begin{equation}}
\def\ee{\end{equation}}
\def\bea{\begin{eqnarray}}
\def\eea{\end{eqnarray}}

\begin{document}
\vspace*{4cm}
\title{Inflation with large gravitational waves.}

\author{ A. Vikman}

\address{ASC, Physics Department LMU, Theresienstr. 37, Munich, Germany}

\maketitle\abstracts{
It is well known that in manifestly Lorentz invariant theories with
nontrivial kinetic terms, perturbations around some classical
backgrounds can travel faster than light. These exotic "supersonic" models
may have interesting consequences for cosmology and astrophysics.
In particular, one can show \cite{my} that in such theories the contribution 
of the gravitational waves to the CMB fluctuations can be significantly
larger than that in standard inflationary models. This increase of 
the tensor-to-scalar perturbation ratio leads to a larger
B-component of the CMB polarization, thus making the prospects for
future detection much more promising. Interestingly, the spectral 
index of scalar perturbations and mass of the scalar field considered in the model 
are practically indistinguishable from the standard case. 
Whereas the energy scale of inflation and hence the reheating temperature 
can be much higher compared to a simple chaotic inflation.}

\section{Introduction}

One of the main consequences of inflation is the generation of primordial
cosmological perturbations \cite{mc} and the production of long wavelength
gravitational waves (tensor perturbations)~\cite{star}. The predicted
slightly red-tilted spectrum of the scalar perturbations is at present in
excellent agreement with the measurements of the CMB fluctuations \cite{cmb}. 
The observation of primordial gravitational waves together with the detection 
of a small deviation of the spectrum from flat would give us further strong 
confirmation of inflationary paradigm. The
detection of primordial gravitational waves is not easy, but they can
be seen indirectly in the B-mode of the CMB polarization (see,
for example, \cite{book}). In standard slow-roll inflationary scenarios 
\cite{Chaot} the amplitude of the tensor perturbations can, in principle, be
large enough to be observed. 
However, it is only on the border of detectability in future experiments.

There are a lot of inflationary scenarios where
the tensor component produced during inflation is much less 
then that in the chaotic inflation. In particular, in
models such as new inflation~\cite{New} and hybrid inflation~\cite{Hybrid},
tensor perturbations are typically small~\cite{book}. Moreover, in the
curvaton scenario \cite{curva} and k-inflation \cite{k-Inflation}, they can
be suppressed completely.

A natural question is whether the gravitational waves can be
significantly enhanced compared to standard scanarious. Recently it was
argued that the contribution of tensor perturbations to the CMB anisotropy
can be much greater than expected \cite{Bartolo:2005jg,Sloth:2005yx}.
However, it was found in \cite{lms} that in the models considered in 
\cite{Bartolo:2005jg,Sloth:2005yx} one cannot avoid the production of too large
scalar perturbations and therefore they are in contradiction with
observations.

In the paper \cite{my} we introduced a class of 
inflationary models where the B-mode of polarization can  
exceed that predicted by simple chaotic inflation. 
These models resemble both k-inflation \cite{k-Inflation}  
and chaotic inflation \cite{Chaot}. Inflation occurs due to the 
potential term in the Lagrangian, and the kinetic term 
has a nontrivial structure responsible for the large sound speed 
of perturbations. In this talk I will review the model from \cite{my}.

\section{Basic equations and main idea}
The generic action describing a scalar field interacting with the
gravitational field is
\begin{equation}
S=S_{g}+S_{\varphi }=\int d^{4}x\sqrt{-g}\left[ -\frac{R}{16\pi }+p(\phi ,X)%
\right] ,  \label{1}
\end{equation}
where $R$ is the Ricci scalar and $p(\phi ,X)$ is a function of the scalar
field $\phi $ and its first derivatives 
$X=\frac{1}{2}\nabla _{\mu }\phi \nabla ^{\mu }\phi$.
We use Planck units, where $G=\hbar =c=1.$ In the case of the usual scalar
field the $X-$dependence of $p$ is trivial, namely, $p=X-V\left( \phi
\right) ,$ while k-inflation and k-essence \cite{k-Inflation,k-Essence,book} 
are based on the non-trivial dependence of $p$
on $X.$ For $X>0$, variation of the action (\ref{1}) with respect to the metric
gives the energy momentum tensor for the scalar field in the form of an
``hydrodynamical fluid'': 
\begin{equation}
T_{~\nu }^{\mu }=(\varepsilon +p)u^{\mu }u_{\nu }-p\delta _{\nu }^{\mu }.
\label{2}
\end{equation}
Here the Lagrangian $p(\phi ,X)$ plays the role of pressure,
the \textquotedblleft four-velocity\textquotedblright\ is 
$u_{\mu }=\nabla _{\mu }\phi/\sqrt{2X}.$ and the energy
density is given by $\varepsilon=2Xp_{,X}-p$ where 
$p_{,X}=\partial p/\partial X$.
Let us consider a spatially flat Friedmann universe with small perturbations:
\begin{equation}
ds^{2}=\left( 1+2\Phi \right) dt^{2}-a^{2}(t)\,\left[ \left( 1-2\Phi \right)
\delta _{ik}+h_{ik}\right] dx^{i}dx^{k},  \label{4}
\end{equation}%
where $\Phi $ is the gravitational potential characterizing scalar metric
perturbations and $h_{ik}$ is a traceless, transverse perturbations
describing the gravitational waves. The minimal set of equations for the 
evolution of the scale factor $a\left( t\right) $ and 
the scalar field $\phi\left( t\right)$ is given by
\begin{equation}
H^{2}\equiv \left( \frac{\dot{a}}{a}\right) ^{2}=\frac{8\pi }{3}\varepsilon~~
~~\textrm{and}~~~~\ddot{\phi}+3c_{S}^{2}H\dot{\phi}+\frac{\varepsilon _{,\phi }}{\varepsilon
_{,X}}=0,
\label{fe}
\end{equation}%
where the dot denotes the derivative with respect to time $t$ and 
the \textquotedblleft speed of sound\textquotedblright\ is 
\begin{equation}
c_{S}^{2}\equiv \frac{p_{,X}}{\varepsilon _{,X}}=\left[ 1+2X\frac{p_{,XX}}{%
p_{,X}}\right] ^{-1} . \label{ss}
\end{equation}%
One can show that $c_{S}$ is in fact the speed of propagation of the 
cosmological perturbations \cite{mg,book}. 
The stability condition with respect to the high frequency cosmological 
perturbations requires $c_{S}^{2}>0$.
For simplicity let us consider theories with Lagrangians of the form 
$p=K(X)-V(\phi )$.
From the equations of motion (\ref{fe}) it is clear that, if the 
slow-roll conditions 
\begin{equation}
XK_{,X}\ll V,~~\textrm{and}~~K\ll V,~~~~\left\vert \ddot{\phi}
\right\vert \ll \frac{V_{,\phi }}{\varepsilon _{,X}}
\end{equation}
are satisfied for at least 75 e-folds then we have a successful slow-roll
inflation due to the potential $V.$ In contrast to ordinary
slow-roll inflation one can arrange here practically any speed of sound $%
c_{S}^{2}$ by taking an appropriate kinetic term $K\left( X\right) $ \cite{my}. 
The crucial point is that the amplitude of the final
scalar perturbations (during the postinflationary, radiation-dominated
epoch) and the ratio of tensor to scalar amplitudes on supercurvature 
scales are given by (see, \cite{book}): 
\begin{equation}
\delta _{\Phi }^{2}\simeq \frac{64}{81}\left( \frac{\varepsilon }{%
c_{S}\left( 1+p/\varepsilon \right) }\right) _{c_{S}k\simeq Ha},~~~~~~
\frac{\delta _{h}^{2}}{\delta _{\Phi }^{2}}\simeq 27\left( c_{S}\left( 1+%
\frac{p}{\varepsilon }\right) \right) _{k\simeq Ha}.  \label{scp}
\end{equation}

Here it is worthwhile reminding that all physical quantities on the right hand 
side of Eqs. (\ref{scp})  have to be calculated during inflation 
at the moment when perturbations with wave number $k$ cross 
corresponding Horizon: $c_{S}k\simeq Ha$ for $\delta _{\Phi }$ and  $k\simeq Ha$ for 
$\delta _{h}$ respectively.   
The amplitude of the scalar perturbations $%
\delta _{\Phi }$ is a free parameter of the theory which is taken to fit the
observations. Therefore, it follows from (\ref{scp}) that the
tensor-to-scalar ratio can be arbitrarily enhanced in such models.

\section{``Simple'' model}
As a concrete example let us consider a simple model with Lagrangian 
\begin{equation}
p(\phi ,X)=\alpha ^{2}\left[ \sqrt{1+\frac{2X}{\alpha ^{2}}}-1\right] -\frac{%
1}{2}m^{2}\phi ^{2},  \label{sm1}
\end{equation}%
where constant $\alpha $ is a free parameter. For $2X\ll \alpha ^{2}$ one
recovers the Lagrangian for the usual free scalar field. The function $p$ is
a monotonically growing concave function of $X$, therefore the system 
is ghost-free. The effective speed of sound,
$c_{S}^{2}=1+2X/\alpha ^{2}$, is larger than the speed of 
light, approaching it as $X\rightarrow 0$. In the slow-roll regime and for $p$ given in (\ref{sm1}), equations (\ref{fe}) reduce to 
\begin{equation}
H\simeq \sqrt{\frac{4\pi }{3}}m\phi,~~~~~3p_{,X}H\dot{\phi}%
+m^{2}\phi \simeq 0.  \label{sm5}
\end{equation}%
For $12\pi\alpha^{2}>m^{2}$ there exists a slow-roll solution:
\begin{equation}
\dot{\phi}\simeq -\frac{mc_{\star}}{\sqrt{12\pi }},~~~~\textrm{where}~~~~
c_{\star }=\left( 1-\frac{m^{2}}{12\pi \alpha ^{2}}\right) ^{-1/2},  \label{sm8}
\end{equation}%
is the sound speed during inflation. The sound speed 
is constant and can be arbitrarily large, 
if we take $12\pi \alpha^{2}\rightarrow m^{2}$.
The pressure and energy density during the slow-roll regime are given by 
\begin{equation}
p\simeq m^{2}\left( \frac{1}{12\pi }\frac{c_{\star }^{2}}{1+c_{\star }}-%
\frac{\phi ^{2}}{2}\right) ,\text{ \ }\varepsilon \simeq m^{2}\left( \frac{1%
}{12\pi }\frac{c_{\star }}{1+c_{\star }}+\frac{\phi ^{2}}{2}\text{\ }\right)
,  \label{sm9}
\end{equation}
respectively. And for the scale factor we have 
\begin{equation}
a\left( \phi \right) \simeq a_{f}\exp \left( \frac{2\pi }{c_{\star }}\left(
\phi _{f}^{2}-\phi ^{2}\right) \right) ,  \label{sm12}
\end{equation}   
where we have introduced subscript $f$ for the quantities at the end of inflation.
The inflation is over when 
$\left(\varepsilon +p\right)/\varepsilon\simeq c_{\star}/\left(6\pi \phi ^{2}\right)$  becomes of order unity, that is, at $\phi \sim \phi_{f}=\sqrt{c_{\star}/6\pi }$. 
After that the field $\phi $ begins to oscillate and decays. Given a number of
e-folds before the end of inflation $N,$ we find that at this time 
$2\pi \phi ^{2}/c_{\star}\simeq N$, and, hence, 
$\left( \varepsilon +p \right) / \varepsilon \simeq 1/3N$
does not depend on $c_{\star}.$ Thus, for a given scale, which crosses the
Hubble scale $N$ e-folds before the end of inflation, the tensor-to-scalar
ratio is 
$\delta _{h}^{2}/ \delta _{\Phi }^{2}\simeq 27c_{\star }\left( 1+
p/\varepsilon \right) \simeq 9c_{\star}/N.$
It is clear that by choosing $\alpha $ close to the critical value $m/\sqrt{%
12\pi }$ we can have a very large $c_{\star}$ and consequently enhance this
ratio almost arbitrarily. Finally 
one can estimate the mas which is needed in order to reproduce the observed 
$\delta _{\Phi } \sim 10^{-5}$. Combining estimations made above we obtain 
$m \simeq 3 \sqrt{3 \pi } \delta _{\Phi }/4N$ or for $N \sim60$, $m\sim3\cdot10^{-7}$ 
similarly to the usual chaotic inflation. The spectral index of scalar perturbations 
reads \cite{book} : 
$n_{s}-1 \simeq -3 \left(1+p/ \varepsilon \right)
-H^{-1}d\left(\ln \left(1+p/ \varepsilon \right)\right)/dt \simeq -2/N$ this is 
exactly the same tilt as for the usual chaotic inflation.  

\section{Conclusions}

We have shown above that in theories where the Lagrangian is a
nontrivial, nonlinear function of the kinetic term, the scale of inflation
can be pushed to a very high energies without coming into conflict 
with observations.
As a result, the amount of produced gravitational waves can be much
larger than is usually expected. If such a situation were realized in
nature then the prospects for the future detection of the B-mode of CMB
polarization are greatly improved. Of course, the theories where this happens
are very unusual. For example, the Cauchy problem is well-posed not for
all initial data \cite{rendall,adams}. 
Moreover, the horizons lose their universality \cite{bmv,SS}.   
Therefore, future observations of
the CMB fluctuations are extremely important since they will not only restrict
the number of possible candidates for the inflaton but also shed light on the problem 
of the ``superluminal'' propagation.

\section*{Acknowledgments}
I am grateful to Viatcheslav Mukhanov and Eugeny Babichev for very useful 
discussions and fruitful collaboration.

\end{document}